\documentclass[prd,a4paper,superscriptaddress,nofootinbib,10pt]{revtex4} 
\usepackage{graphicx}
\usepackage{amssymb}
\usepackage{amsmath}
\usepackage{lscape}
\usepackage{mathrsfs}
\usepackage{textcomp}
\usepackage{epsfig}
\usepackage{slashed}

\usepackage{lineno,hyperref}
\usepackage{amsmath,amsfonts,amssymb}
\modulolinenumbers[5]

\newcommand{\be}{\begin{equation}}
\newcommand{\ee}{\end{equation}}
\newcommand{\bea}{\begin{eqnarray}}
\newcommand{\eea}{\end{eqnarray}}

\begin{document}
\title{Conformal quantum mechanics and holography in noncommutative space-time}

\author{Kumar S. Gupta}
\email{kumars.gupta@saha.ac.in}
\affiliation{Theory Division, Saha Institute of Nuclear Physics, 1/AF Bidhannagar, Kolkata 700064, India}

\author{E. Harikumar}
\email{harisp@uohyd.ernet.in}

\author{Zuhair N. S.}
\email{zuhairns@gmail.com}
\affiliation{School of Physics, University of Hyderabad, Central University P.O.,
Hyderabad 500046, India}

\date{\today}

\begin{abstract}

We analyze the effects of noncommutativity in conformal quantum mechanics (CQM) using the $\kappa$-deformed space-time as a prototype. Upto the first order in the deformation parameter, the symmetry structure of the CQM algebra is preserved but the coupling in a canonical model of the CQM gets deformed. We show that the boundary conditions that ensure a unitary time evolution in the noncommutative CQM can break the scale invariance, leading to a quantum mechanical scaling anomaly. We calculate the scaling dimensions of the two and three point functions in the noncommutative CQM which are shown to be deformed.  The $AdS_2/CFT_1$ duality for the CQM suggests that the corresponding correlation functions in the holographic duals are modified. In addition, the Breitenlohner-Freedman bound also picks up a noncommutative correction.  The strongly attractive regime of a canonical model of the CQM exhibit quantum instability. We show that the noncommutativity softens this singular behaviour and its implications for the corresponding holographic duals are discussed.

\end{abstract}

\maketitle










\section{Introduction}

Conformal invariance is a fundamental symmetry of physics, which is typical for massless systems that have no dimensionful parameters. For such systems, the Poincare symmetry gets enlarged to include dilatation and special conformal transformations. A simple yet rich class of systems which capture the essential features of this enhanced symmetry is described by conformal quantum mechanics (CQM) \cite{fubini}. The action of a system described by CQM is invariant under the full conformal group, whose algebra is given by $so(2,1)$. A diverse range of physical phenomena can be analyzed within the framework of CQM, which include $AdS_2$/$CFT_1$ duality \cite{maldacena,maldacena1,jackiw}, near-horizon conformal structure of black holes \cite{carlip,danny1,danny2}, Calogero type models \cite{biru1,biru2,pijush,andjelo}, scaling anomalies \cite{esteve,camblong}, renormalization in quantum mechanics \cite{kumar1}, scattering of electrons from polar molecules \cite{pulak,mol2}, effects of Coulomb charge and topological defects in graphene \cite{castro,sen1,sen2,sen3} and study of the zeros of the Riemann Zeta function \cite{amilcar,oka}.

Given the relevance of CQM to  black hole physics \cite{carlip,danny1,danny2} or holography \cite{maldacena,maldacena1,jackiw}, it is pertinent to ask about the behaviour of CQM at very high energies such as the Planck scale. There are several candidates for describing Planck scale physics including string theory, loop gravity and noncommutative geometry. In this paper we shall take noncommutative geometry as a framework for describing physics at the Planck scale and study CQM in that context. It is known that the quantum uncertainty principle and Einstein's general relativity together lead to a  general class of noncommutative spacetimes \cite{dop1,dop2}.  For our analysis, we consider the $\kappa$-deformed algebra \cite{luk1,luk2,luk3} as a prototype of the noncommutative spacetime, which is relevant for a variety of black holes \cite{brian1,brian2,ohl,ncbh1,ncbh2,ncbh3}. The $\kappa$-deformed algebra admits a large class of realizations \cite{meljanac1,meljanac2,meljanac3}. Here we work with a specific subclass that is adapted to the CQM. In addition, keeping in mind the weakness of the Planck scale effects, we keep terms only upto the first order in the noncommutative deformation parameter. Under these assumptions the noncommutativity preserves the symmetry structure of the CQM but the dimensionless parameter in a canonical model of the CQM undergo rescaling. Our analysis applies to the conformal world line symmetries, but not to the target space conformal symmetries, which would be broken by $\kappa$-deformations.

The study of how scale and conformal invariance is broken in physical systems is often as important as the analysis of the symmetry itself. In particular, how a system with no dimensionful parameter can give rise to bound states with finite energies has been investigated in a variety of physical systems. In order for the bound state to exist in such a system, the scaling symmetry must be broken \cite{biru1,esteve,jackiw-beg}. The source of this anomaly can often be traced to the boundary conditions, which can either break the scale invariance \cite{biru1} or prevent $so(2,1)$ from being the spectrum generating algebra \cite{biru2}. The choice of appropriate boundary conditions is determined by the requirement of a unitary time evolution, which demands that the corresponding Hamiltonian be self-adjoint \cite{reed}. Thus the self-adjoint extensions \cite{reed} determine the boundary conditions that can lead to a quantum breakdown of the scale invariance \cite{biru1,jackiw-beg}.

In this work we shall assume that the quantum dynamics governed by the noncommutative CQM respects unitarity \cite{baluni} and the corresponding Hamiltonian must therefore be self-adjoint \cite{reed}. This leads to a special class of boundary conditions which break the scale invariance, resulting in a quantum mechanical scaling anomaly \cite{biru1,esteve,camblong}. One might wonder if there would be any scale invariance in the noncommutative framework at all, as the deformation parameter introduces a length scale. We shall show that upto the first order in the deformation parameter, the symmetry structure of the CQM is preserved and the noncommutativity itself does not destroy the scale invariance upto this order. On the other hand, the coupling appearing in the Hamiltonian of the canonical model of the CQM \cite{fubini} gets deformed, whose consequences are explored in our work. 

There exists an interesting connection between the CQM and the $AdS_2/CFT_1$ duality \cite{maldacena,maldacena1,jackiw}. It is well known that the killing vectors of the $AdS_2$ span a $so(2,1)$ algebra, which is also the symmetry algebra of the CQM. In this work we have explored certain aspects of this holographic duality within the noncommutative framework. The concept of a noncommutative holography was discussed in \cite{manin} and further explored in the context of noncommutative black holes \cite{tajron1,tajron2,tajron3}. In the present analysis, we calculate the correlation functions in the noncommutative CQM which are shown to be deformed. Using the $AdS_2/CFT_1$ duality, we argue that the scaling dimensions of operators in the holographic dual are also deformed. In the gravity side, the constraints on the scaling dimensions lead to the well known Breitenlohner-Freedman bound \cite{bf1,bf2}. We show that the Breitenlohner-Freedman bound acquires a noncommutative correction which has the proper commutative limit. 

In the strongly attractive regime of the canonical CQM Hamiltonian, there is a breakdown of the quantum vacuum \cite{landau,case}. A renormalization group approach has been developed to address this problem \cite{jackiw-beg,kumar1}. Here we show that upto the first order in the deformation parameter the quantum breakdown of the vacuum  is softened  by noncommutativity, but not completely removed. Using holographic ideas, we relate the renormalization group flow of the coupling to that of the Breitenlohner-Freedman bound. This has interesting implications for the holographic dual, which requires further analysis.

This letter is organized as follows. In Section 2, we study the conformal invariance of inverse square potential problem in $\kappa$-deformed spacetime. In Section 3 the effect of noncommutativity on the scaling anomaly is discussed.  Next in Section 4 we discuss the renormalization of the strongly attractive inverse square interaction in the presence of noncommutativity. Section 5 discusses the $AdS_2/CFT_1$ holography and the effect of noncommutativity on the associated physical quantities. Section 6 is devoted to conclusions and discussions.

\section{Conformal quantum mechanics in $\kappa$-spacetime}

Consider the $so(2,1)$ Lie algebra generated by three generators $H$, $D$ and $K$ satisfying the relations 
\be 
\label{alg}
[H,D] = i\hbar H, \quad [K,D] =-i\hbar K, \quad [H,K] = 2i\hbar D.  
\ee
This algebra can be realized by a large class of operators. For the analysis in this paper, we shall work with a canonical model of the CQM, given by the operators
\be
\label{trio}
H = \frac{1}{2} \left ( p^2 + \frac{\alpha}{q^2} \right ), ~~~~~~~ 
D = -\frac{1}{4} \left ( p~q + q~p \right ), ~~~~~~~
K = \frac{1}{2} q^2 
\ee
where $p$ and $q$ are canonically conjugate  variables satisfying the relation
\be
\left [ q,p \right ] = i\hbar.
\ee
The constant $\alpha$ is dimensionless and taken to be real. The generator $H$ corresponds to a particle of unit mass under the action of an inverse-square potential. The generators $D$ and $K$ correspond to dilatation and special conformal transformations respectively.

Our goal is to study the conformal quantum mechanics described by operators in (\ref{trio}) in noncommutative $\kappa$-deformed space-time.  For a general introduction to quantum mechanics in $d$-dimensional $\kappa$-deformed spacetime, see \cite{luk}. The $\kappa$-deformed space-time is defined by the algebra,
\be
\label{kappa} 
[\hat{x}_0, \hat{x}_1] = ia \hat{x}_1,
\ee
with  $a = \frac{1}{\kappa}$ being the deformation parameter \cite{luk1,luk2,luk3}. We shall take the parameter $a$ to be real and positive. We are taking $x_0=ct$, where $c$ is the velocity of light. With this convention, $a$ has the dimension of $[L]$ and $\kappa$ has dimension of $1\over L$. It may be mentioned that in natural units, where $\hbar=c=1$, $\kappa$ would have mass dimension. Note that in our approach $a$ or $\kappa$ is assumed to be a fundamental constant. Note that since the origin of the noncommutativity is supposed to be due to the Planck scale effects, it may be possible to identify $a$ with the Planck length $l_P \equiv \sqrt{\frac{\hbar G}{c^3}}$. Instead of working with operators as in (\ref{kappa}), it is convenient to work with commutative functions and derivatives which realize the same operators. The $\kappa$-space admits a large class of realizations \cite{meljanac1,meljanac2,meljanac3} and a choice has to be made which is adapted to the problem under consideration. In the subsequent analysis we shall work with a particular realisation of $\kappa$-space-time in which the coordinates are given by \cite{meljanac1,meljanac2,meljanac3}
\be 
\hat{x}_1 = x_1 \varphi(a), \quad \hat{x}_0 = x_0,  \label{xrealiz}
\ee
where $\varphi$ is arbitrary function whose form depends on the choice of realization. 

We will now obtain the Hamiltonian for a free particle in $\kappa$-deformed space-time using the expression for dispersion relation in $1+1$ dimensional space-time, in a specific choice of realization (For a details on study of $\kappa$-spacetime, see \cite{lowdim1,lowdim2,lowdim3,lowdim4}). Let $\hat{p}_\mu = -i\hbar D_\mu$ be the momentum in $\kappa$-spacetime, with $D_\mu$ being the generalised derivative having the following form in our realization \cite{meljanac1,meljanac2,meljanac3}, $ D_1 = \partial_1 \frac{e^{-A}}{\varphi},\quad D_0 = \partial_0 \frac{\sinh A}{A} - ia \partial_1^{2} \frac{e^{-A}}{2\varphi^{2}},$ with $A = ia \partial_0 = a\frac{p^{0}}{\hbar}$.  We write the dispersion relation $\hat{p}_\mu \hat{p}^{\mu} + m^{2} = 0$  and use $\hat{p}_{\mu} = -i \hbar D_{\mu}$ (with metric signature $(-+)$) to obtain
\bea 
\frac{4 \hbar^{2}}{a^{2}} \sinh^{2} \left ( \frac{ap^{0}}{2\hbar} \right ) - p_1^{2} \frac{e^{-\frac{ap^{0}}{\hbar}}}{\varphi^{2}\left(\frac{ap^{0}}{\hbar}\right)} + \frac{a^{2}}{4 \hbar^{2}} \left[ \frac{4 \hbar^{2}}{a^{2}} \sinh^{2}\left(\frac{ap^{0}}{2\hbar}\right) -  p_1^{2} \frac{e^{-\frac{ap^{0}}{\hbar}}}{\varphi^{2}(\frac{ap^{0}}{\hbar})} \right]^{2} = m^{2}c^{2}. \label{disp}
\eea
Note that $p^0$ and $p^1=p_1$ are the components of the momenta in the commutative space, which are related to the corresponding derivatives using $\hbar$ \cite{meljanac1,meljanac2,meljanac3}. Thus the process of constructing the realizations introduces $\hbar$, which subsequently appears in the dispersion relation. We also note that each term of Eqn. (6) has the dimension of the square of the momentum. It may be mentioned that if we work in natural units with $\hbar=c=1$, then $\kappa$ has dimension $[M]$ and $a$ has dimension of $1\over{[M]}$. In that case Eqn. (6) has the form
\bea 
4 \kappa^2 \sinh^{2} \left ( \frac{p^{0}}{2\kappa} \right ) - p_1^{2} \frac{e^{-\frac{p^{0}}{\kappa}}}{\varphi^{2}\left(\frac{p^{0}}{\kappa}\right)} + \frac{1}{4 \kappa^2 } \left [4 \kappa^2 \sinh^{2}\left(\frac{p^{0}}{2\kappa}\right) -  p_1^{2} \frac{e^{-\frac{p^{0}}{\kappa}}}{\varphi^{2}(\frac{p^{0}}{\kappa})} \right]^{2} = m^{2}. 
\eea
Again each term on both sides of Eqn (7) have dimension $[M]^2$ and it is dimensionally consistent.

Expanding the above dispersion relation in powers of a, for the choice of $\varphi = e^{-\frac{a p^{0}}{\hbar}}$ and solving for the energy $p_0 $, valid upto first order in $a$, we obtain 
\be 
p^{0} =\frac{E}{c}= \sqrt{m^2 c^{2}+p_1^2} + \frac{1}{2}\frac{ a}{\hbar} p_1^2
\ee
From now on wards, we will denote $p_1^{2}$ as $p^{2}$, keeping in mind that we are working in $1+1$ dimensional spacetime.
 In the non-relativistic limit, this leads to the free particle Hamiltonian, 
\be 
H_0 = \frac{p^{2}}{2m}(1+\frac{amc}{\hbar}) \equiv \frac{p^{2}}{2m} f(a) \label{ho},
\ee
 where $f(a)= (1+a\nu_c)$ and $\nu_c = \frac{mc}{\hbar}$ is the inverse Compton wavelength of the particle of mass $m$.  
 
 Here we have considered a particle of general mass $m$ and in our later calculations, we will be setting mass to be unity, as was chosen for the Hamiltonian in Eqn. (\ref{trio}).  We will now write the potential, $V(\hat{r}) = \frac{\alpha}{\hat{r}^{2}}$ using the realization in \eqref{xrealiz} and we find $V(\hat{r})= \frac{\alpha}{r^{2}}+ O(a^{2})$.  

Thus, the $\kappa$ space-time Hamiltonian $H$, upto first order in $a$, for a particle with mass, $m=1$ moving in inverse square potential is
\be
H = \frac{1}{2} \left ( f(a)~ p^2 + \frac{\alpha}{r^2} \right ).  \label{ham}
\ee
From now onwards, we work with units (unless otherwise mentioned) in which $\hbar = 1$ and $m = 1$.  We re-express the above Hamiltonian as
\be 
\label{htilde}
\tilde{H} \equiv \frac{H}{f(a)} = \frac{1}{2} \left ( p^2 + \frac{\tilde{\alpha}}{r^2} \right ), ~~~~~~~~~ 
\tilde{\alpha}(a) \equiv \frac{\alpha}{f(a)}. 
\ee
Note that the factor arising from noncommutativity is a dimensionless quantity and hence, the rescaled coupling $\tilde{\alpha}$ is dimensionless as well. Even though parameter $a$ is a dimensionful quantity, the effect of $\kappa$-deformation appears only through a dimensionless combination. For this reason, we expect the Hamiltonian to obey conformal symmetry, as shown below. With this rescaled Hamiltonian, we obtain the algebra for the system as
\be 
\label{alg1}
[\tilde{H},\tilde{D}] = i \tilde{H}, \quad [\tilde{K},\tilde{D}] =-i \tilde{K}, \quad [\tilde{H},\tilde{K}] = 2i \tilde{D},
\ee
where the generators $\tilde{D} = D$ and $\tilde{K}=K$ and thus do not have any dependence on the deformation parameter $a$.

We have thus seen that upto the first order in the deformation parameter, the effect of the noncommutativity in the conformal quantum mechanics is captured by the rescaling of the coupling $\alpha$ in the generator $H$, which appears as the Hamiltonian for a large class of physical systems. Note also that, the $p$ appearing in the Hamiltonian in $\kappa$-spacetime given in Eqn. \eqref{htilde} is the usual momentum operator. Thus the only parameter in the theory will be evolution parameter $t$. Thus the Hamiltonian used in this paper study the conformal field theory in $0+1$ dimension. 

It may be noted that the $\kappa$-deformed space-time under consideration admits a class of twist operators that satisfy the co-cycle and other conditions of a Hopf algebra, leading to the twisted co-products for symmetry of the system in the co-algebra sector. For the particular choice of realization considered here, the twisted co-product $\Delta$ is given by $\Delta = \mathcal{F}^{-1} \Delta_0 \mathcal{F}$, where $\Delta_0$ is the commutative co-product and the twist operator is given by $\mathcal{F} = e^{-N\otimes A}$ with $N = q \partial_{q}$ and $A = i\frac{a}{c}\partial_{t}$. For details see \cite{meljanac1,meljanac2,meljanac3}. 

We end this section with a comment on the effect of the noncommutativity on the spectrum of the $so(2,1)$ algebra. The generators of $so(2,1)$ algebra in the Cartan basis can be written as   
\bea 
\tilde{S} &=& \frac{1}{2} \left ( \frac{\tilde{K}}{l} + l \tilde{H} \right ), \ \quad \tilde{J}_{\pm} = \frac{1}{2} \left ( \frac{\tilde{K}}{l}+ l \tilde{H} \right ) \pm i \tilde{D}. \label{tildes}
\eea
They satisfy the algebra
\be 
\label{so(2,1)}
[\tilde{S}, \tilde{J}_{\pm}] = \pm \tilde{J}_{\pm}, \quad [\tilde{J}_-, \tilde{J}_+] = 2 \tilde{S}. 
\ee
The operator $\tilde{S}$ is compact and its spectrum is given by \cite{fubini}
\bea 
\label{ev}
\tilde{S} |m> = s_m |m>;  \quad s_m = s_0 +m, \, m= 0,1,2, \cdots, 
\eea 
where
\be
\label{s0} 
s_0 (a) = \frac{1}{2} \left( 1\pm \sqrt{\tilde{\alpha}(a)+\frac{1}{4}} \right).
\ee
Here we normally choose the positive sign in front of the square root to ensure that $s_0 > 0$ \cite{fubini}. We now express the eigenvector in terms of raising operator $\tilde{J}_+$ as
\be 
\label{vec}
|m> = \sqrt{\frac{\Gamma(2s_0(a))}{m! ~\Gamma(2s_0(a)+m)}} (\tilde{J}_+)^{m} |0>.
\ee
Thus it is clear that the representations of the $so(2,1)$ algebra appearing in the CQM is directly affected by the noncommutative deformation parameter.  We shall next study the effect of this rescaling on the nature of the spectrum of $H$ and how it affects the scale invariance of the system. 

\section {Scaling anomaly}

In this section we shall study certain quantum aspects of the Hamiltonian $\tilde{H}$. To this end, consider the Schrodinger's equation for bound state
\be
\label{sch}
\tilde{H} \psi = -E \psi .
\ee
 We can re-write this equation as
\be
\label{eff}
H_e \psi \equiv \left [ -\frac{d^2}{dr^2} + \frac{\tilde{\alpha}}{r^2} \right ] \psi = -2 E \psi,
\ee
where $r \in (0, \infty)$. Eqn. (\ref{eff})  corresponds to the Schrodinger's equation for an effective Hamiltonian $H_e = 2 {\tilde{H}}$ and henceforth we shall work with $H_e$.  

In order to analyze the Schrodinger's equation (\ref{eff}), we need to specify the boundary conditions. In this process our guiding principle would be unitarity, which is assumed to be valid at the Planck scale where noncommutative effects are relevant \cite{baluni}. The requirement of unitarity demands that the boundary conditions be so chosen that the operator $H_e$ would be self-adjoint \cite{reed}.  The conditions for self-adjointness of an operator of the type of $H_e$ has been discussed in the literature \cite{biru1}. It is known that $H_e$ is a symmetric or Hermitian operator \cite{reed} in the domain $\mathcal{D}_0(H_e) \equiv \{\phi (0) = \phi^{\prime} (0) = 0,~ \phi,~ \phi^{\prime}~  {\rm absolutely~ continuous} \} $, but is not necessarily self-adjoint in it. In fact, when $-\frac{1}{4} \leq \tilde{\alpha} < \frac{3}{4}$, the operator $H_e$ admits a one-parameter family of self-adjoint extensions labelled by a parameter $z \in R$ mod ($2 \pi$) \cite{biru1}. For each value of $\tilde{\alpha} \in (-\frac{1}{4}, \frac{3}{4})$ and for a generic value of $z$, the operator $H_e$ admits a single bound state with energy given by \cite{biru1}
 \be
E  
 = - \frac{1}{2} \left [ \frac{{\rm sin}(\frac{z}{2} + 3 \pi \frac{\sqrt{{\tilde{\alpha}} + \frac{1}{4}}}{4})}
{{\rm sin}(\frac{z}{2} + \pi \frac{\sqrt{{\tilde{\alpha}} + \frac{1}{4}}}{4})} \right ]^{\frac{1}{\sqrt{{\tilde{\alpha}} + \frac{1}{4}}}} 
\ee
and for ${\tilde{\alpha}} = \frac{1}{4}$, the corresponding bound state energy is given by \cite{biru1}
\be
E=  
- {\rm {exp}} \left [ \frac{\pi}{2}  {\rm {cot}} 
\frac{z}{2} \right ]. \label{h1}
\ee

The appearance of the bound state may seem surprising since the scale invariant operator $H_e$ does not contain any dimensionful parameter and thus should not admit a bound state. The reason why the bound state appears in its spectrum is related to the fact that when $-\frac{1}{4} \leq \tilde{\alpha} < \frac{3}{4}$, the domain or equivalently the boundary conditions which render $H_e$ self-adjoint break the scale invariance. For the parameter range 
$-\frac{1}{4} \leq \tilde{\alpha} < \frac{3}{4}$, the domain $\mathcal{D}_z(H_e)$ in which $H_e$ is self-adjoint is different from $\mathcal{D}_0(H_e)$. It can be shown that the dilatation operator does not leave the domain $\mathcal{D}_z(H_e)$ invariant. In other words, if  $\psi \in \mathcal{D}_z(H_e)$, then for a generic value of the self-adjoint extension parameter $z$, the dilatation operator $D$ acts in such a way that $ D \psi \notin \mathcal{D}_z(H_e)$ \cite{biru1}. This is an example of scaling anomaly, whereby the classical scale invariance is broken through the choice of boundary conditions, which is a purely quantum effect.

Let us now address the question as to how noncommutativity affects the breakdown of scale invariance. Recall that the effective coupling 
$\tilde{\alpha}$ in the noncommutative case is related to the commutative coupling $\alpha$ as $\tilde{\alpha} = \frac{\alpha}{f(a)}$. Thus the range $-\frac{1}{4} \leq \tilde{\alpha} < \frac{3}{4}$ corresponds to $-\frac{1}{4} f(a) \leq \alpha < \frac{3}{4} f(a)$.
Now consider a range of the commutative coupling $\alpha \in [ -\frac{f(a)}{4}, \frac{3 f(a)}{4})$, where $f(a) > 1$. It is clear that 
for certain parts of this range for the commutative coupling $\alpha$, namely between $[-\frac{f(a)}{4}, -\frac{1}{4})$ and between  $ (\frac{3}{4}, 
\frac{3 f(a)}{4})$, there would be no self-adjoint extension and hence no violation of the scaling symmetry due to quantum 
effects. Now once the noncommutativity is turned on, the entire range $ [ -\frac{f(a)}{4}, \frac{3 f(a)}{4})$ gets mapped to 
$ [ -\frac{1}{4}, \frac{3}{4})$, for which, as we have argued before, the scaling anomaly exists. Thus we can conclude that for certain 
ranges of coupling which in the commutative case do not show scaling anomaly, would do so once the noncommutativity is turned on.

\section{Renormalization}

In this section we shall consider the case when $\tilde{\alpha} < - \frac{1}{4}$, in which case we can write $\tilde{\alpha} = -\nu^2 - \frac{1}{4}$, $\nu \in R$. The equation (\ref{eff}) can be written as
\be
\label{ren} 
\left [ \frac{d^{2}}{dr^{2}} + \frac{\nu^2 + \frac{1}{4}}{4} - 2E \right ] \psi = 0.
\ee
It is well known that the energy spectrum obtained from this equation is unbounded from below, signalling the onset of the famous ``the fall to the centre'' problem as originally discussed by Landau \cite{landau,case}. The fact that the energy is unbounded from below corresponds to the breakdown of the quantum vacuum. The range of coupling for which this happens is called the supercritical range, $-\frac{1}{4}$ being the critical value of the coupling.

In order to understand how the singularity appears, we introduce an ultraviolet regulator in the position space at $r = b$ and solve the Eqn. (\ref{ren}) with the boundary condition $\psi(r=b) = 0$ \cite{kumar1}. The general solutions of Eqn. (\ref{ren}) is given by 
\be 
\psi(x) = \sqrt{x} K_{i\nu} (x); \quad x= -i \sqrt{2E}r,
\ee
where with $K_{i\nu}$ is the modified Bessel function of the third kind with imaginary order \cite{abr}. The zeroes of this function appear at 
\be 
x_n = e^{-n\pi /\nu} (2 e^{-\gamma}) (1+O(\nu)); \quad \text{where $\gamma$ is the Euler number,}
\ee
which leads to the energy spectrum given by
\be 
E_n = -  e^{-2n\pi /\nu} \frac{(2 e^{-\gamma})^{2}}{2 b^{2}} (1+O(\nu)).
\ee
where $n = 1, 2, ...\infty$. The ground state energy corresponds to $n=1$ and  $n= \infty$  is an accumulation point of the spectrum. As discussed above, the ground state energy becomes unbounded from below as the ultraviolet regulator $b \rightarrow 0$. 

We now use the technique of renormalization \cite{kumar1} to address this divergence. As a first step, we make the coupling $\nu$ a function of the cutoff $b$. Next we demand that the ground state energy becomes independent of the cutoff as the ultraviolet regulator is removed. This is enforced by imposing the condition
\be
\lim_{b \rightarrow 0} \frac{dE_1}{db} =0.
\ee
This condition leads to the $\beta$-function given by
\be
\label{beta} 
\beta(b) = - b \frac{d\nu(b)}{db} = -\frac{\nu^{2}}{\pi} + {\text{subleading terms}}.
\ee
From Eqn. (\ref{beta}) it is clear that the $\beta$-function admits an ultraviolet stable fixed point at $\nu =0$ or equivalently at $\tilde{\alpha} = -\frac{1}{4}$. 

Recalling again the relation $\tilde{\alpha} = \frac{\alpha}{f(a)}$, we see that if we started with a commutative coupling such that $\alpha = -\frac{f(a)}{4}$, then the corresponding noncommutative coupling is  $\tilde{\alpha} = -\frac{1}{4}$. Thus with $f(a) > 1$, a coupling which was leading to a ultraviolet divergence in the commutative theory can still lead to a proper spectrum in the noncommutative version. The effect of the noncommutativity is thus to soften the divergence in the sense described above.

\section{Correlation functions and $AdS_2$/$CFT_1$ correspondence}

In this section we shall discuss the effect of noncommutativity on the correlation functions that appear in the canonical model of the CQM. Note that the algebra generated by the killing vector fields of $AdS_2$ also span $so(2,1)$, which has led to the well known $AdS_2/CFT_1$ correspondence \cite{maldacena1,jackiw}. We shall see the implications of this correspondence on the noncommutative CQM.

Recall that the Hamiltonian $\tilde{H}$ as one of the generators of the $so(2,1)$ algebra and that it is related to time-translation invariance. Let $\tau$ be the parameter which describes the evolution of system. We introduce the vectors in the $\tau$-basis as $|\tau>$. In this basis, Hamiltonian have the representation $H = i \frac{d}{d\tau}$. Following the method of \cite{fubini}, the $\kappa$-deformed two-point function is found to have the form
\be 
A_2(\tau _2,\tau _1) \equiv < \tau _1| \tau _2> = \sum_m \psi^{*}_m(\tau _1) \psi_m (\tau _2) = \Gamma(2s_0(a)) \left(\frac{1}{2i (\tau _1 - \tau _2)^{2s_0(a)}}\right). \label{A2}
\ee
It is interesting to note that the effective dimension, $s_0(a)$ has a dependence on the deformation parameter. Similarly, the $\kappa$-deformed three-point is found to be
\bea
A_3(\tau _1,\tau _2,\tau _3) &=& <\tau _1| B(t) |\tau _2>
  \propto   \, \frac{1}{|\tau -\tau _1|^{\delta} |\tau _2-\tau |^{\delta}|\tau _1-\tau _2|^{-\delta +2s_0(a)}}, \label{A3}
\eea
where $\delta$ is the dimension of the operator $B$, which is otherwise arbitrary.

The structure of the two and three point functions obtained above are exactly of the form of two and three point functions in $AdS_2$ 
\cite{adsguide}, which are given by
\bea 
G_2(x,y) &=& <O(x) O(y)>  \propto \, \frac{1}{|x-y|^{\Delta}} \\
G_3(x, y ,z) &=& <O(x) O(y)O(z)> \propto   \, \frac{1}{|x-y|^{\Delta_1 +\Delta_2 - \Delta_3} |y-z|^{\Delta_2 +\Delta_3 - \Delta_1}|z-x|^{\Delta_3 +\Delta_1 - \Delta_2}}.
\eea
We have here, considered arbitrary operators  $O(x_j)$  with scaling dimension, $\Delta_j$. The identification, $A_3 \sim G_3$ implies that two of the operators ($O(x)$ and $O(z)$) have the scaling dimension $\Delta = s_0(a)$ and third operator ($O(y)$) have the dimension $\delta$. It is interesting to observe that the effect of noncommutativity appears through the modification of the dimension of the operators. We emphasize that the dimension have an explicit dependence on the deformation parameter $a$ and also, the dimension attains its value in commutative case when the deformation parameter is set to zero.

The dependence of the effective dimension on non-commutative parameter $a$ has some interesting implications. In order to understand this, let us choose a specific choice of realization, say $\varphi = e^{-ap^{0}}$. This leads to expression for $s_0$ as
\be 
s_0 = \frac{1}{2} \left( 1+ \sqrt{\frac{\alpha}{1+a\nu_c} +\frac{1}{4}} \right). \label{r0f1}
\ee
This tells us that the effect of non-commutativity on the conformal dimension $s_0(a)$ of the operators in the two point function can be physically related to a shift of the coupling strength. Since $s_0$ is real, we obtain
\be 
\label{bf}
\alpha \geq -\frac{1}{4} (1+a\nu_c).
\ee
Thus we see that the reality of $s_0$, which appears in the eigenvalues of the generator $\tilde{S}$ in Eqn (\ref{ev}), leads to the constraint on the coupling $\alpha$, which depends on the noncommutative deformation parameter. 

The $AdS_2/CFT_1$ correspondence \cite{maldacena1} relates the mass of a scalar field in $AdS_2$ to the scaling dimension of the corresponding operators in the CFT. It is well known that the mass of the scalar field in $AdS_2$ is constrained by the Breitenlohner-Freedman bound. In the commutative case, the corresponding bound on the CFT side leads to the constraint $\alpha \geq -\frac{1}{4}$. The result in Eqn. (\ref{bf}) gives a noncommutative correction to the constraint on $\alpha$. Using the $AdS_2/CFT_1$ correspondence we can now argue that the  Eqn. (\ref{bf}) produces a noncommutative correction to  the Breitenlohner-Freedman bound. Note that the modified Breitenlohner-Freedman bound corresponds to the lower bound on the coupling $\alpha$ arising from the self-adjoint extension of $H$. We shall make further remarks on this in the concluding section.

\section{Conclusion}

In this work we have reformulated a canonical model of the CQM in the $\kappa$-Minkowski space-time. The noncommutativity encodes the features of physics  at the Planck scale \cite{dop1,dop2}. Since observable effects of such features are expected to be weak, we have analyzed the noncommutative effects only upto the first order in the deformation parameter. In the chosen canonical model of the CQM \cite{fubini}, the Hamiltonian $H$ contains a dimensionless parameter $\alpha$, which appears as a coupling of the inverse square potential. The effect of non-commutativity is introduced into the Hamiltonian through the factor $f(a)$ (see eqn.\eqref{htilde}).  Here it is to be emphasized that the deformation parameter appears through a dimensionless combination of $\hbar$,$c$ and $m$.  

In order to arrive at this Hamiltonian, we first obtained the Hamiltonian for non-relativistic free particle in $\kappa$-spacetime from the $\kappa$-deformed dispersion relation. Further, since the only parameter that is present in our theory is the evolution parameter $t$, we conclude that we are
studying quantum mechanics (i.e., field theory in $0+1$ spacetime). It is clear that the quantity $f(a)$, which contains the mass of the particle and deformation parameter $a$, is dimensionless. Since the $\kappa$-deformation appears only through this combination, which is dimensionless, non-relativistic conformal invariance is shown to be valid in $\kappa$-spacetime. In the presence of the $\kappa$-deformed noncommutativity, the algebraic structure of the CQM is preserved upto the first order in the deformation parameter. The main effect of the noncommutativity is to scale the dimensionless coupling $\alpha$ appearing in $H$, while the form of the other two generators of $so(2,1)$ remain unchanged. The eigenvalues and the eigenvectors of the $so(2,1)$ algebra are directly affected by the noncommutativity. In particular, the quadratic Casimir of the $so(2,1)$ algebra and the scaling dimensions appearing in the correlation functions are also modified.

If we assume that the CQM describes a unitary time evolution, then the Hamiltonian operator $H$ must be self-adjoint. For a range of the coupling $\alpha$, the requirement of unitarity leads to boundary conditions that violates the scaling symmetry. For these class of boundary conditions, the generator of dilatation does not leave the domain of $H$ invariant, which leads to a scaling anomaly \cite{biru1}. This effect persists in the presence of noncommutativity and the system continues to admit a scaling anomaly for a certain range of the coupling. In particular, a certain range of the coupling which in the commutative theory would preserve the scale invariance now leads to a scaling anomaly once the noncommutativity is turned on. Thus the length scale associated with the noncommutativity clearly affects the scaling symmetry of the CQM, but does not completely destroy it, at least upto the first order in the deformation parameter.

In a recent analysis, the $AdS_2/CFT_1$ duality has been used to put bounds on the scaling dimension of the CQM vacuum \cite{okazaki}. The scaling dimension is related to the quadratic Casimir of the $so(2,1)$, which depends on $\alpha$. Thus the bound on $\alpha$ arising from that on the scaling dimension is due to the $AdS_2/CFT_1$ correspondence. Recall that purely within the CQM, there is another bound on $\alpha$ arising from the requirement of a unitary time evolution, or equivalently from the self-adjoint extensions of $H$. What we have found is that these two bounds arising from apparently different origins coincide exactly. The lower limit of the constraint on $\alpha$ is related to the Breitenlohner-Freedman bound, which relates the unitarity of the CQM to the stability of the $AdS_2$. Moreover, our analysis shows that the Breitenlohner-Freedman bound is modified due to noncommutativity. The upper limit of the constraint is also affected by noncommutativity, whose interpretation in terms of the gravity dual is not completely clear and is a subject of ongoing investigation. 

It is known that when the inverse-square coupling in $H$ is strongly attractive, there is a quantum breakdown of the vacuum. We have shown that a certain range of the coupling which was supercritical in the commutative theory, becomes subcritical in the presence of the noncommutativity. In this sense, the effect of the noncommutativity is to soften the singularity in the supercritical sector, although the singularity is not completely removed. Note that the critical value of the coupling corresponds to the Breitenlohner-Freedman bound. From both the gravity side and the CFT side we know that there is a singularity beyond the critical value. Here we have presented a renormalization group analysis which addresses the singularity problem in the CFT side. Using the $AdS_2/CFT_1$ correspondence we can expect that there would be a similar renormalization group analysis in the gravity side as well.

\section*{Acknowledgement}
ZNS would like to acknowledge the support for this work received from CSIR, Govt. of India under CSIR-SRF scheme.
EH thanks SERB, Govt. of India, for support through EMR/2015/000622.


\section*{References}

\begin{thebibliography}{50}
\bibitem{fubini} Vittorio de Alfaro, S. Fubini and G. Furlan, Nuovo Cim. {\bf A 34} (1976) 569.
\bibitem{maldacena} J. Maldacena, Adv. Theor. Math. Phys. 2 (1998) 231, arXiv:hep-th/9711200.
\bibitem{maldacena1} J. Maldacena and D. Stanford, Phys.Rev. {\bf D94} (2016) 106002, arXiv:hep-th/1604.07818.
\bibitem{jackiw} C. Chamon, R. Jackiw, S. Pi and L. Santos, Phys. Lett. {\bf B 701} (2011) 503.
\bibitem{carlip} S. Carlip, {\it Quantum gravity in 2+1 dimensions}, Cambridge University Press, 2003.
\bibitem{danny1} D. Birmingham, Kumar S. Gupta and S. Sen, Phys. Lett. {\bf B505}, 191 (2001).
\bibitem{danny2} Kumar S. Gupta and  S. Sen, Phys. Lett. {\bf B526}, 121 (2002).	
\bibitem{biru1} B. Basu-Mallick, Pijush K. Ghosh and Kumar S. Gupta, Nucl. Phys. {\bf B659}, 437 (2003). 
\bibitem{biru2} B. Basu-Mallick, Pijush K. Ghosh and Kumar S. Gupta, Phys. Lett. {\bf A311}, 87 (2003).
\bibitem{pijush} Pijush K. Ghosh and Kumar S. Gupta, Phys. Lett. {\bf A323}, 29 (2004).
\bibitem{andjelo} B. Basu-Mallick, Kumar S. Gupta, S. Meljanac and A. Samsarov, Eur. Phys. J. {\bf C 58}, 159 (2008).
\bibitem{esteve} J. G. Esteve, Phys. Rev. {\bf D34}, 674 (1986).
\bibitem{camblong} H. E. Camblong and C. R. Ordonez, Phys. Rev. {\bf D68} (2003) 125013.
\bibitem{kumar1} Kumar S. Gupta and S.G. Rajeev, Phys. Rev. {\bf D 48} (1998) 5940
\bibitem{pulak} P. R. Giri, Kumar S. Gupta, S. Meljanac and A. Samsarov, Phys. Lett. {\bf A 372}, 2967 (2008).
\bibitem{mol2} H. E. Camblong, L. N. Epele, H. Fanchiotti, C. A. Garcia Canal, Phys. Rev. Lett. {\bf 87} (2001) 220402.
\bibitem{castro} Valeri N. Kotov, Bruno Uchoa, Vitor M. Pereira, F. Guinea, and A. H. Castro Neto, Rev. Mod. Phys. {\bf 84}, 1067 (2012).
\bibitem{sen1} Kumar S. Gupta and Siddhartha Sen, Mod. Phys. Lett. {\bf A 24}, 99 (2009).
\bibitem{sen2} Kumar S. Gupta and Siddhartha Sen, Phys. Rev. {\bf B 78}, 205429 (2008).
\bibitem{sen3} Kumar S. Gupta and Siddhartha Sen, J. Phys. A: Math. Theor. {\bf 46} 055303 (2013).
\bibitem{amilcar} Kumar S. Gupta, E. Harikumar and Amilcar R. de Queiroz, EPL {\bf 102}, 10006 (2013).
\bibitem{oka} T. Okazaki, Phys. Rev. {\bf D92}, 126010 (2015).
\bibitem{dop1} S. Doplicher, K. Fredenhagen and J. E. Roberts, Phys. Lett. {\bf B 331}, 39 (1994).
\bibitem{dop2} S. Doplicher, K. Fredenhagen and J. E. Roberts,  Comm. Math. Phys. {\bf 172}, 187 (1995).
\bibitem{luk1} J. Lukierski, A. Nowicki, H. Ruegg and V. N. Tolstoy, Phys. Lett. {\bf B 264}, 331 (1991).
\bibitem{luk2} J. Lukierski, A. Nowicki and H. Ruegg, Phys. Lett. {\bf B 293}, 344 (1992).
\bibitem{luk3} J. Lukierski and H. Ruegg, Phys. Lett. {\bf B 329}, 189 (1994).
\bibitem{brian1} B. P. Dolan, Kumar S. Gupta and A. Stern, Class. Quantum Grav. {\bf 24}, 1647 (2007).
\bibitem{brian2} B. P. Dolan, Kumar S. Gupta and A. Stern, J. Phys. Conf. Ser.  {\bf 174}, 012023 (2009).
\bibitem{ohl} T. Ohl and A. Schenkel, JHEP {\bf 0910}, 052 (2009).
\bibitem{ncbh1} 
  K.~S.~Gupta, S.~Meljanac and A.~Samsarov,
  Phys.\ Rev.\ D {\bf 85} (2012) 045029.
  
\bibitem{ncbh2} Kumar S. Gupta, S. Meljanac and A. Samsarov, Phys.Rev. {\bf D 85}, 045029 (2012).
\bibitem{ncbh3} K.~S.~Gupta, E.~Harikumar, T.~Juric, S.~Meljanac and A.~Samsarov,
  Adv.\ High Energy Phys.\  {\bf 2014} (2014) 139172.
\bibitem{meljanac1}  S. Meljanac and M. Stojic, Eur. Phys. J. {\bf C 47} (2006)  531.
\bibitem{meljanac2}  S. Meljanac, A. Samsarov, M. Stojic and Kumar S. Gupta, Eur. Phys. J. {\bf C 53} (2008) 295.
\bibitem{meljanac3} T. R. Govindarajan, Kumar S. Gupta, E. Harikumar, S. Meljanac and D. Meljanac, Phys. Rev. {\bf D 77}, 105010 (2008) , arXiv:0802.1576 [hep-th].
\bibitem{jackiw-beg} R. Jackiw in M. A. B. Beg Memorial Volume, edited by A. Ali and P. Hoodbhoy, World Scientific, Singapore, 1991.
\bibitem{reed} M. Reed and B. Simon, {\it Methods of Modern Mathematical Physics}, volume 2, Academic Press, New York, 1972.
\bibitem{baluni} A.P. Balachandran, T.R. Govindarajan, C. Molina and P. Teotonio-Sobrinho, JHEP {\bf 0410} (2004) 072.
  Phys.\ Rev.\ D {\bf 77} (2008) 105032.
\bibitem{manin} Y. I. Manin and M. Marcolli, Adv. Theor. Math. Phys. {\bf 5} (2002) 617.
\bibitem{tajron1} Kumar S. Gupta, E. Harikumar, Tajron Juric, Stjepan Meljanac and Andjelo Samsarov  Adv. High Energy Phys. {\bf 2014} (2014) 139172.
\bibitem{tajron2} Kumar S. Gupta, E. Harikumar, Tajron Juric, Stjepan Meljanac, Andjelo Samsarov, JHEP {\bf 1509}  (2015)025.
\bibitem{tajron3} Kumar S. Gupta, Tajron Juric, Andjelo Samsarov, arXiv:1703.00514 [hep-th].
\bibitem{bf1} P. Breitenlohner and D. Z. Freedman, Phys. Lett. {\bf B 115}, 197 (1982).
\bibitem{bf2} P. Breitenlohner and D. Z. Freedman, Ann. Phys. {\bf 144}, 249 (1982).
\bibitem{landau} L. D. Landau and E. M. Lifshitz, {\it Quantum Mechanics, Pergamon}, London, 1958.
\bibitem{case} K. M. Case, Phys. Rev. 80, 797 (1950).
\bibitem{luk} J. Lukierski, H. Ruegg and W. J. Zakrzewski, Annals Phys. {\bf 243} (1995) 90.
\bibitem{lowdim1} F. Cianfrani, J. Kowalski-Glikman, D. Pranzetti and G. Rosati, Phys.Rev. {\bf D94} (2016) 084044.
\bibitem{lowdim3} A. Pachoł and P. Vitale, J.Phys. {\bf A48} (2015) 445202.
\bibitem{lowdim2} A. Borowiec, Kumar S. Gupta, S. Meljanac and A. Pachol, Europhys.Lett. {\bf 92} (2010) 20006.
\bibitem{lowdim4}  P. Aschieri, A. Borowiec and A. Pachol, arXiv:hep-th/1703.08726.
\bibitem{abr} I. S. Gradshteyn and I. M. Ryzhik, {\it  Table of Integrals, Series, and Products}, Academic Press, Seventh edition, 2007.
\bibitem{adsguide}  M. Natsuume, {\it AdS/CFT Duality User Guide}, Springer, 2015. 
\bibitem{okazaki} T. Okazaki, ArXiv: hep-th/1704.00286 (2017).
\end{thebibliography}
\end{document}